\documentclass[runningheads]{llncs}
\usepackage[T1]{fontenc}
\usepackage{graphicx}
\usepackage{amsmath}
\usepackage{booktabs}%
\usepackage{hyperref}
\usepackage{subcaption}
\begin{document}
\title{Multi Scale Curriculum CNN for Context-Aware Breast MRI Malignancy Classification}
\titlerunning{Curriculum CNN for Breast MRI Malignancy Classification}
% If the paper title is too long for the running head, you can set
% an abbreviated paper title here
%
\author{Christoph~Haarburger$^1$, Michael~Baumgartner$^1$, Daniel~Truhn$^{2,1}$, Mirjam~Broeckmann$^2$, Hannah~Schneider$^2$, Simone~Schrading$^2$, Christiane~Kuhl$^2$, Dorit~Merhof$^1$}
% \author{*\\
% *\\
% *}
% %
\authorrunning{C. Haarburger et al.}
% % First names are abbreviated in the running head.
% % If there are more than two authors, 'et al.' is used.
% %
\institute{%
$^1$Institute of Imaging and Computer Vision, RWTH Aachen University, Germany\\
$^2$Dpt. of Diagnostic and Interv. Radiology, University Hospital Aachen, Germany\\
\email{christoph.haarburger@lfb.rwth-aachen.de}}
\maketitle              % typeset the header of the contribution
\begin{abstract}
Classification of malignancy for breast cancer and other cancer types is usually tackled as an object detection problem:
Individual lesions are first localized and then classified with respect to malignancy.
However, the drawback of this approach is that abstract features incorporating several lesions and areas that are not labelled as a lesion but contain global medically relevant information are thus disregarded:
especially for dynamic contrast-enhanced breast MRI, criteria such as background parenchymal enhancement and location within the breast are important for diagnosis and cannot be captured by object detection approaches properly.

In this work, we propose a 3D CNN and a multi scale curriculum learning strategy to classify malignancy globally based on an MRI of the whole breast.
Thus, the global context of the whole breast rather than individual lesions is taken into account.
% In a first training stage, localization information on a relatively small dataset is used on 3D patches.
% In the second stage, whole 3D scans and only one global label are used for training.
Our proposed approach does not rely on lesion segmentations, which renders the annotation of training data much more effective than in current object detection approaches.

Achieving an AUROC of 0.89, we compare the performance of our approach to Mask R-CNN and Retina U-Net as well as a radiologist.
Our performance is on par with approaches that, in contrast to our method, rely on pixelwise segmentations of lesions.
\end{abstract}

\section{Introduction}
Detection of anatomical structures or lesions in medical images is mostly solved by object detection algorithms that rely on pixelwise segmentations of individual objects.
% This approach has several drawbacks
% NMS, WBC nötig...
However, there is not always a clear consensus among clinicians on which object is considered a suspicious lesion that should be segmented.
Moreover, segmentations are subject to high inter-rater variability~\cite{menze_2015}.
Both of these aspects limit the quality of training data and a reliable and meaningful evaluation.
The degree of natural variation  cannot be captured by interpreting tumors as objects with hard boundaries.
The latter does make sense in classical computer vision, but for biomedical applications, a more holistic approach that takes the global \textit{context} of a lesion into account is crucial.
Another limitation of the detection and instance segmentation approach lies in the high cost associated with labeling large datasets.
As a consequence, most datasets are rather small, preventing that deep learning algorithms unfold their full potential and limiting the power of evaluation results.
Finally yet importantly, clinicians demand a decision support at patient-level rather than a classification of individual objects/regions of interest to efficiently integrate computer-aided diagnosis algorithms into clinical workflow.\newline
\noindent\textbf{Related Work.}
A naive approach to incorporating global context would be a classification of whole axial slices instead of detecting individual objects.
However, this approach has produced poor results so far because it is a needle-in-haystack kind of problem~\cite{maier-hein_2018}.
Other works have adapted object detection algorithms such as Mask R-CNN~\cite{he_2017} and RetinaNet~\cite{lin_2017} for lesion detection and classification~\cite{jung_2018}. %evtl. noch anfügen: with limited success. Falls wir dazu sinnvoll zitieren können
In~\cite{jaeger_2017}, diffusion-weighted breast MR images are classified for malignancy by setting all voxels outside  the lesion segmentation to zero.
Lotter et al.~\cite{lotter_2017} proposed a curriculum learning strategy for mammogram classification that concatenates features maps from patch level for a global breast malignancy classification.
In a similar approach Koshravan et al.~\cite{khosravan_2018} propose a detection and malignancy classification algorithm for lung nodules by superimposing a grid on CT scans and by classifying each grid cell as benign or malignant.
Maicas et al.~\cite{maicas_2017} proposed a reinforcement learning approach for breast MRI lesion detection using bounding boxes.
Several other authors have proposed methods for lesion classification based on bounding boxes~\cite{amit_2017,dalmis_2018,yan_2018}.
In \cite{maicas_2018}, a curriculum learning strategy is proposed and extended to classification of whole volumes in~\cite{maicas_2018b} showing promising results.
Zhu et al.~\cite{zhu_2017c} propose a multiple instance learning loss for whole mammogram classification.
In \cite{wang_2018a}, a multi scale CNN for detection of lesions in breast ultrasound images is proposed.
Finally Jäger et al.~\cite{jaeger_2018} proposed Retina U-Net, which advances RetinaNet~\cite{lin_2017} by leveraging segmentation supervision for breast lesions.
We introduce a CNN that is trained on several scales in a curriculum learning strategy, which is highly efficient for small datasets.
Moreover, our 3D CNN does not rely on pixelwise segmentations or bounding boxes but rather lesion center points.
This allows for efficient annotation and leverages the whole lesion context for classification at patient level.\newline
\noindent\textbf{Contributions.}
We introduce a simple 3D CNN for breast cancer malignancy classification that does not need to detect lesions individually and does not rely on pixelwise segmentations or bounding boxes.
Moreover, we propose a multi scale curriculum learning strategy that efficiently trains the network on two scales, starting on patch scale and continuing training on the whole breast including all global context.
We provide a comparison with other state of the art methods: Naive whole breast classification, Mask-R-CNN and Retina U-Net.
Finally, in order to maximize reproducibility, comparisons and adaption, we provide all code\footnote{\url{https://github.com/haarburger/multi-scale-curriculum}} including implementations of network architectures, preprocessing and curriculum training in PyTorch.

\section{Image Data}
Our dataset consists of dynamic contrast-enhanced (DCE) MR images of 408 patients from clinical routine at our institution.
Images were acquired on a 1.5\,T Philips Scanner using a standardized protocol consisting of a T2-weighted turbo spin echo sequence and a T1-weighted gradient echo sequence acquired as a dynamic series~\cite{kuhl_2007}.
DCE T1-weighted images were acquired before and every 70\,s after injection of contrast agent (gadobutrol) for four post-contrast agent time points.
The acquisition matrix was \(512 \times 512\), yielding an in-plane resolution of \(0.6\times0.6\)\,mm and slice thickness was 3\,mm.
In order to allow a comparison of our approach with approaches that rely on an auxiliary segmentation such as Mask R-CNN and Retina U-Net, all lesions were manually segmented on every slice by a radiologist with 13 years of experience in breast MRI.

Out of the 408 patients, 305 had malignant and 103 had benign findings.
Malignancy was determined by biopsy; diagnoses of benign lesions were validated by follow up for 24 months.
Since many of the malignant findings were only present in one breast, the overall ratio of malignant and benign samples at breast level (rather than patient level) in the whole dataset is 40.4\%/59.6\%.
All images were resampled to \(512 \times 512 \times 32\) voxels.
No bias field correction was performed.
Finally, we cropped all images by removing voxels that only contain air or tissue from the thorax, leading to images of a spatial resolution of \(512 \times 256 \times 32\) voxels.
\section{Methods}
\subsection{Multi Scale Curriculum Network}

\begin{figure}[htb]
    \centering
    \includegraphics[width=0.99\linewidth]{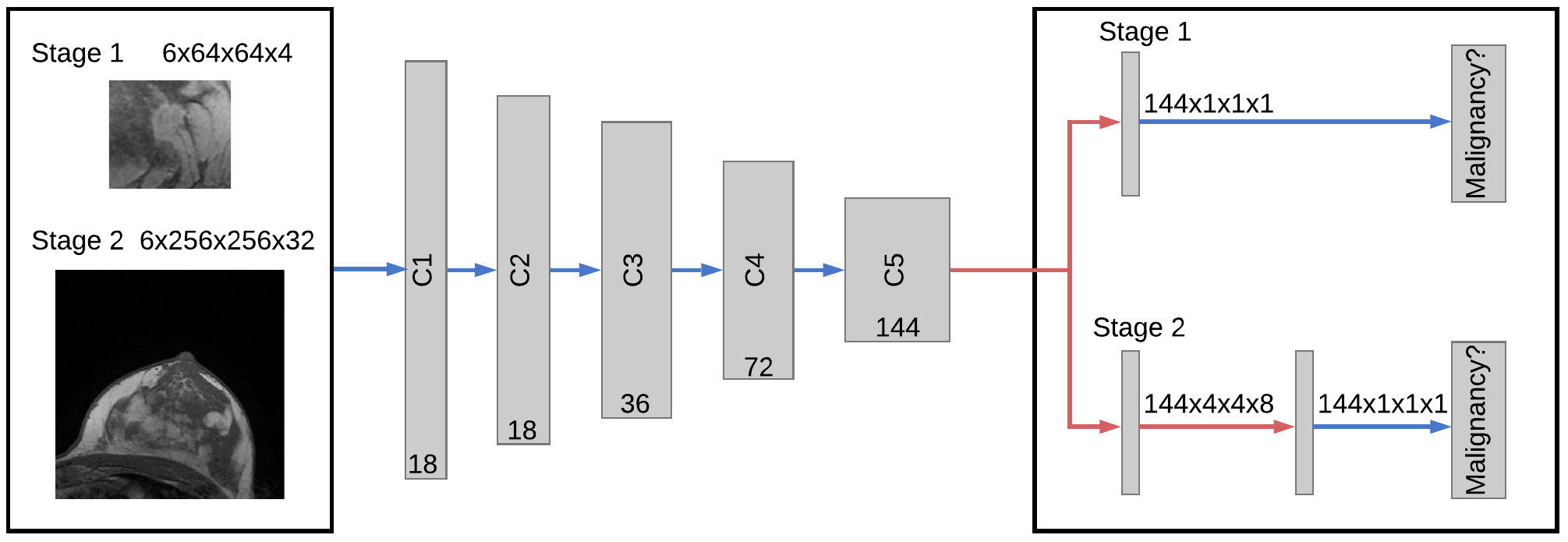}
    \caption{Network architecture: Residual Blocks are named \(C_i\) as in~\cite{he_2016} with channel dimensions indicated at the bottom of each block.
    Convolutions are indicated in blue, pooling operations in red.}
    \label{fig:network}
\end{figure}

The network architecture we propose consists of two basic components as depicted in Fig~\ref{fig:network}:
1) A Backbone network that consists of a 3D CNN to generate feature maps.
2) Classification head that performs a classification based on the aggregated feature maps provided by the two previous components.
For the Backbone we initially evaluated many different achitectures including ResNets~\cite{he_2016} and DenseNets~\cite{huang_2016} of different depths, U-Net~\cite{ronneberger_2015} and Feature Pyramid Network~\cite{lin_2016}.
Since choice of the backbone architecture did not have a significant impact on the overall performance for the final model, we focus on ResNet18 in this work, because of its efficiency.

\noindent\textbf{Multi Scale Curriculum Training.}
Training is conducted in two stages:
\begin{itemize}
    \item Stage 1: Classification of 3D lesion \textit{patches} with size $64\times64\times4$, where each patch contains at least one lesion as determined by the location of the lesion centerpoints.
    If at least one malignant lesion is contained in the patch, the patch label is set to malignant and benign otherwise.
    \item Stage 2: Classification of 3D \text{volumes} containing a whole breast with size $256\times256\times32$.
    To account for the changed input size, the network architecture is modified by introducing an additional adaptive average pooling layer between the original average pooling and fully-connected layer.
    All network parameters are trained during Stage 2.
\end{itemize}
% Mention random offset of patches? Lesions are not always centered for better second stage performance.
% The fully connected layer is replaced too because of a design decision in the early phases of the second stage
From all MRI scans, left and right breast were fed into the network separately.
The five time points from the dynamic T1-weighted series and the T2-weighted series were fed into the CNN in the channel dimension, which leads to an input volume of \(6\times 256\times 256\times 32\) (~channels,~x,~y,~z).
The network was trained with batch size 4, Adam optimizer with default parameters, instance normalization, leaky ReLU activation functions and a learning rate of \(10^{-4}\) and \(10^{-5}\) for Stages 1 and 2, respectively.

For data augmentation, we mirrored all images and rotated around the z axis by \(\pm15^{\circ}\).
Training, hyperparameter tuning and testing were performed in a 5-fold cross validation, using  261 patients (64\%), 65 patients (16\%) and 82 patients (20\%) of the data for training, validation and testing, respectively.
Since each single breast is fed into the network separately, splitting at patient level guarantees that both breasts of a single patient are exclusively contained in only one of the three data subsets.

\subsection{Comparison Methods}
\noindent\textbf{Naive ResNet.}
In this approach, we naively train a vanilla 3D ResNet18 to directly predict malignancy globally without multi scale curriculum learning.
Since only a very small fraction of voxels belongs to a lesion, this approach is a needle-in-a-haystack kind of problem.
It serves as a baseline for the multi scale curriculum learning approach.

\noindent\textbf{Mask R-CNN.}
Mask R-CNN~\cite{he_2017} is a widely used two-stage detection framework that leverages supervision information from a segmentation auxiliary task.

\noindent\textbf{Retina U-Net.}
Based on the RetinaNet one stage detection framework~\cite{lin_2017}, Retina U-Net combines RetinaNet with enhanced segmentation supervision using the U-Net architecture.

\section{Results}
% In order to aggregate the two malignancy scores at patient level, we took the maximum probability of malignancy over the two predictions on breast level.
On a Nvidia GeForce 2080 Ti GPU, training of Stage 1 and 2 takes about 4 and 6 hours, respectively.
Prediction of a single breast can be performed in under 100\,ms.

In order to show the benefit of multi scale curriculum learning, we evaluated Mask R-CNN and Retina U-Net as described above as well as a naive 3D ResNet18 without curriculum training (Stage 2 only) and the proposed approach with multi scale curriculum learning.
In addition, a radiologist with 2 years of experience in breast MRI rated the images with respect to malignancy.
Test performance is provided in Tab.~\ref{tab:results}.
The highest AUROC is achieved by the radiologist, followed by ResNet18 Curriculum and Retina U-Net.
Mask R-CNN performs slighly worse.
The naive 3D image classification approach achieved a very poor performance that is not significantly different from random guessing.
\begin{table}
    \begin{tabular}{@{}lccccc@{}}
    \toprule
                           & AUROC  & Accuracy & \#Parameters \\ \midrule
    Mask R-CNN~\cite{he_2017}       &   0.88 \(\pm\) 0.01   &     0.77 \(\pm\) 0.03     &  3.91M            \\
    Retina U-Net~\cite{jaeger_2018}           &  0.89 \(\pm\) 0.01   &    0.82 \(\pm\) 0.02      &       3.90M       \\
    ResNet18 Naive          &  0.50 \(\pm\) 0.04   &       0.45 \(\pm\) 0.05     &   2.66M           \\
    % ResNet18 MultiScale            &  0.49   &     0.5        &      0.54       &    0.44      &   2.66M           \\
    ResNet18 Curriculum  &   0.89 \(\pm\) 0.01  &     0.81 \(\pm\) 0.02    &          2.66M    \\
    % ResNet18 MultiScale w/ Curriculum &  0.89   &    0.83         &      0.82       &     0.77     &          2.66M    \\
    \hline
    Radiologist            &  0.93   &    0.93      &       -       \\ \bottomrule
    \end{tabular}
    \vspace{0.3cm}
    \centering
    \caption{Test performance of the comparison methods and proposed approach over a 5-fold cross validation.}
    \label{tab:results}
    \end{table}

\begin{figure}
    \centering
    \includegraphics[width=0.65\linewidth]{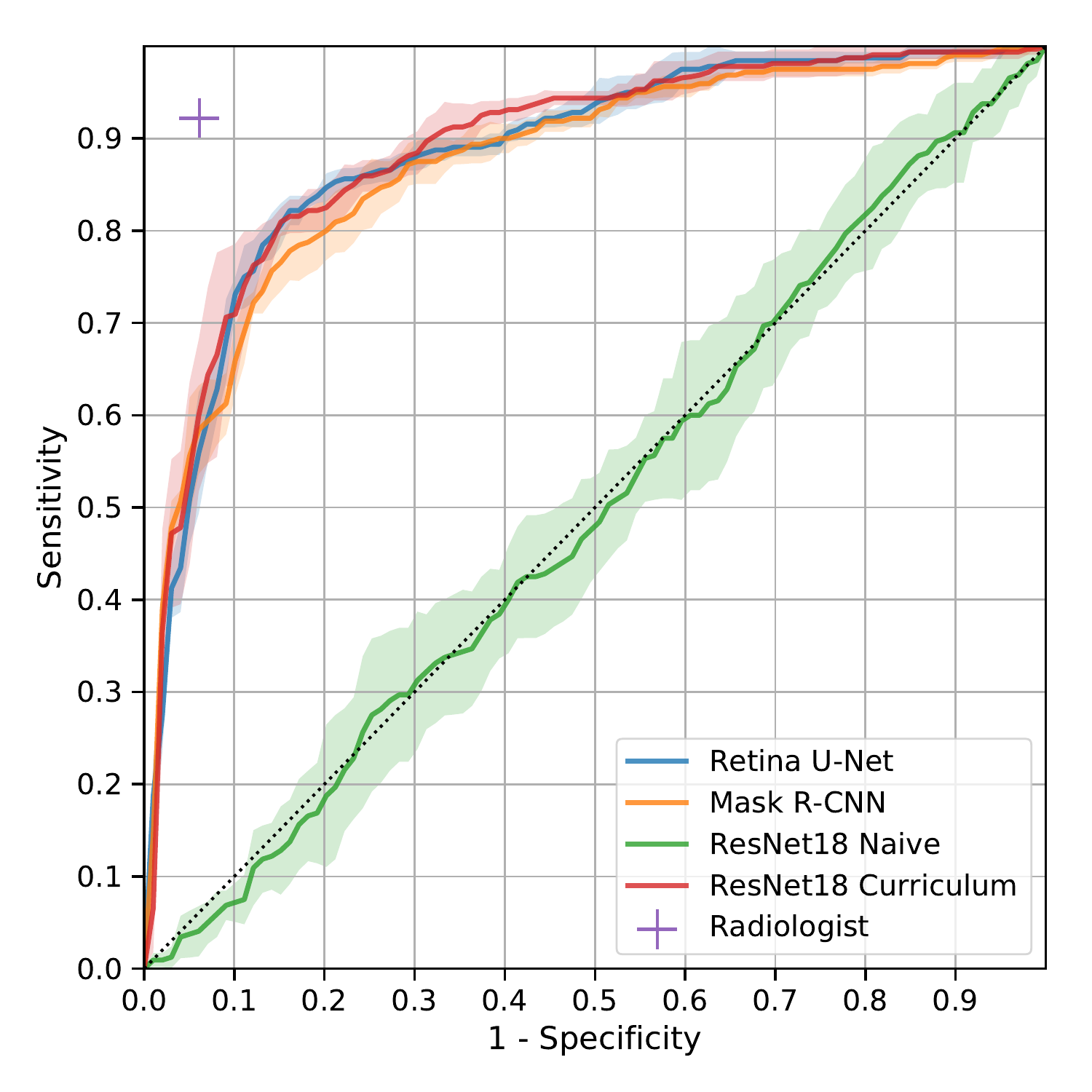}
    \caption{Receiver operating characteristic. Means and standard deviations over the 5-fold cross validation are encoded as lines and contiguous areas, respectively.}
    \label{fig:roc}
\end{figure}

\begin{figure}[htb]
    \centering
    \begin{subfigure}{.16\textheight}
      \centering
      \includegraphics[height=0.15\textheight]{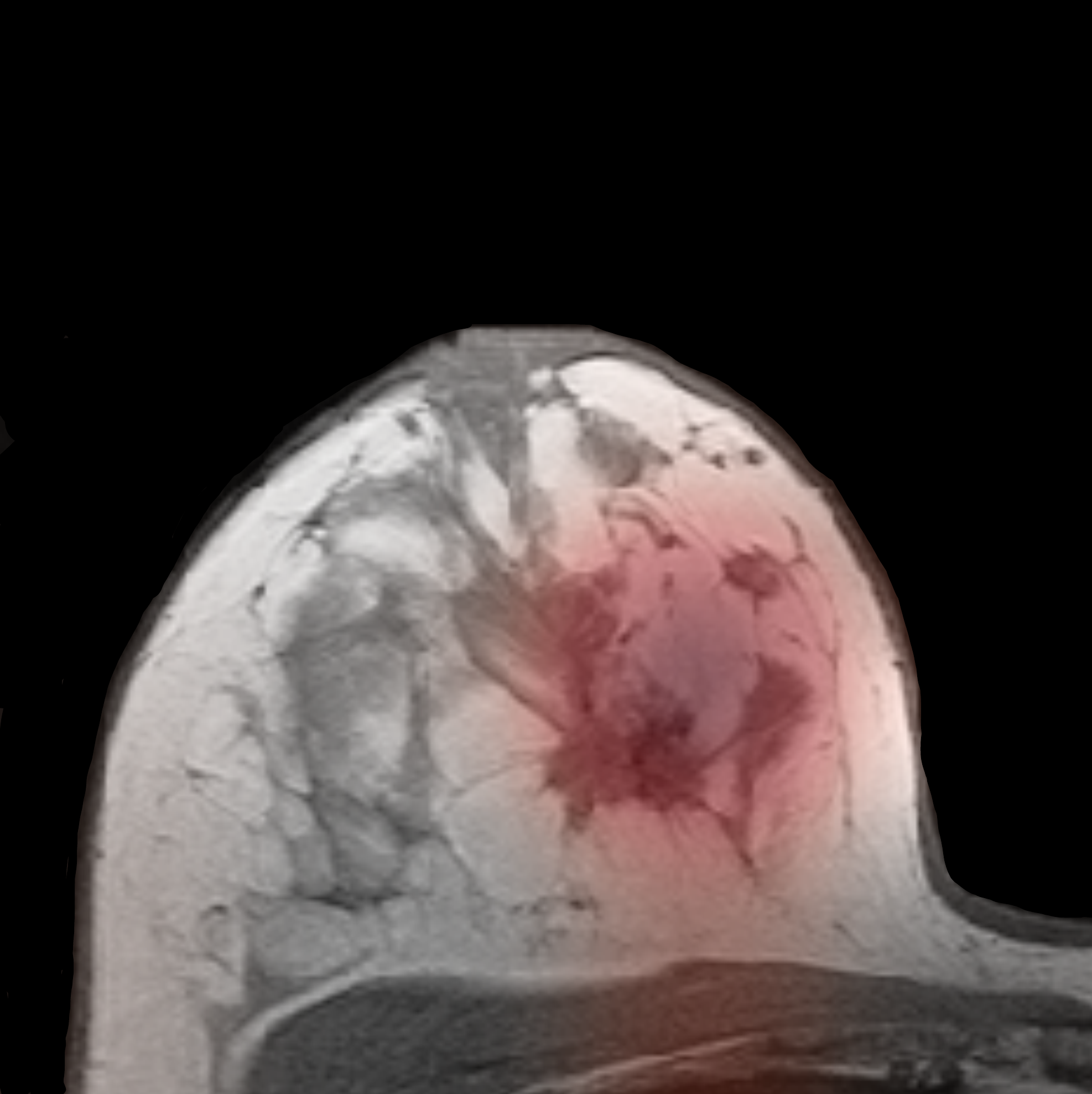}
    %   \caption{Malignant, both correct}
        \caption{}
      \label{fig:sub1}
    \end{subfigure}%
    \begin{subfigure}{.16\textheight}
      \centering
      \includegraphics[height=0.15\textheight]{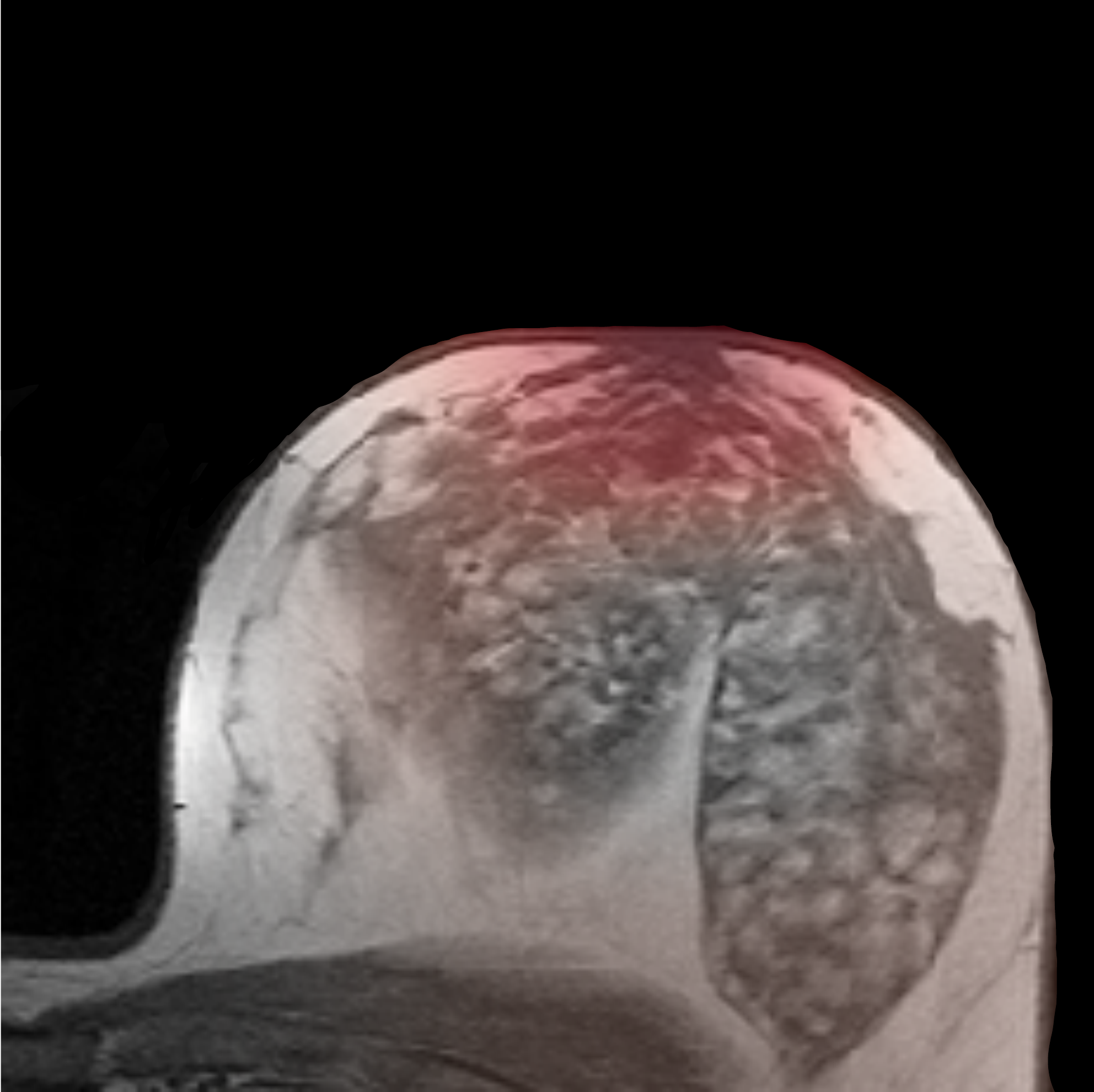}
    %   \caption{Benign, cnn correct, rad incorrect}
      \caption{}
      \label{fig:sub2}
    \end{subfigure}%
    \begin{subfigure}{.16\textheight}
        \centering
        \includegraphics[height=0.15\textheight]{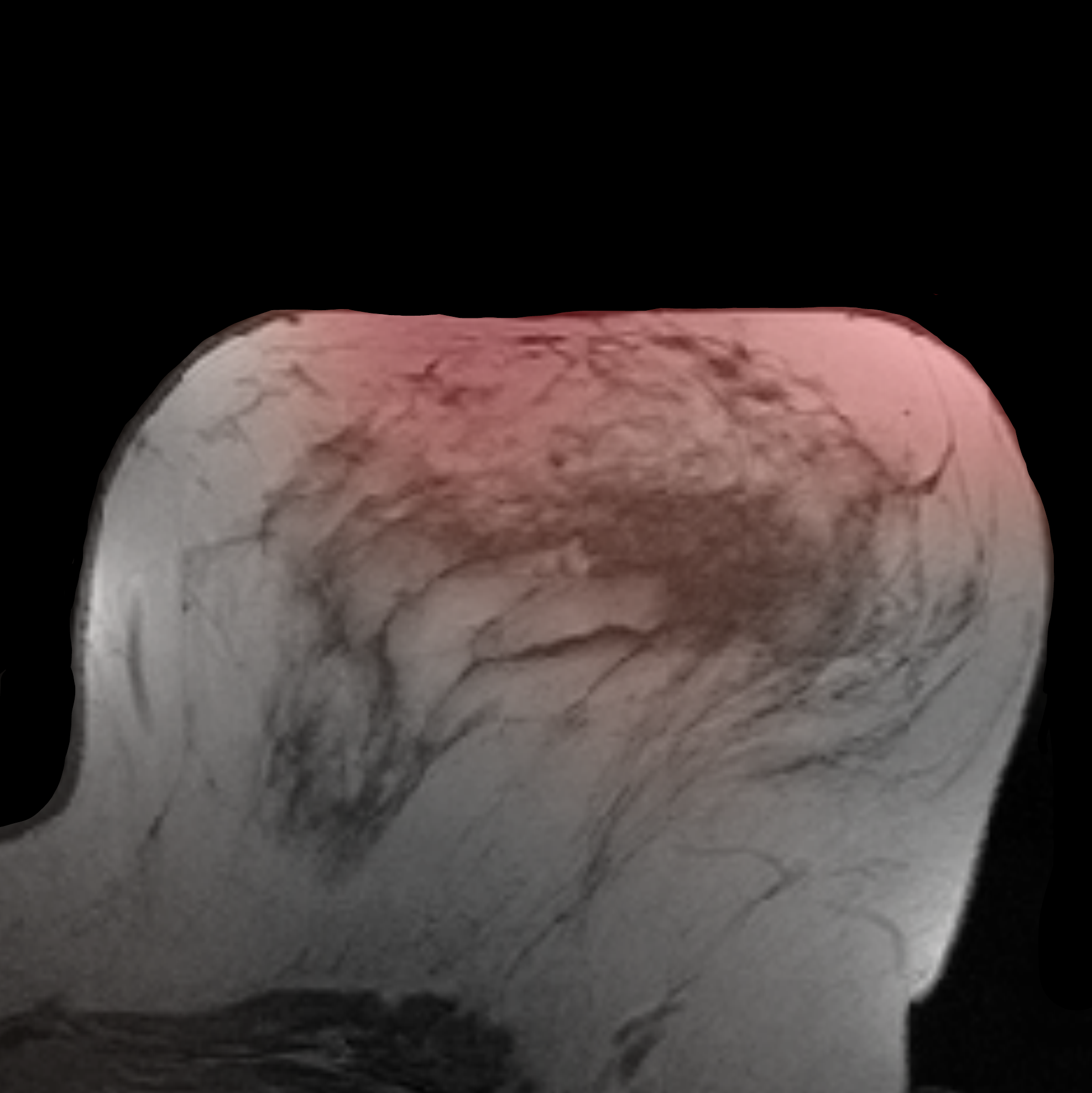}
        % \caption{Malignant, both incorrect}
        \caption{}
        \label{fig:sub2}
      \end{subfigure}%
      \begin{subfigure}{.16\textheight}
        \centering
        \includegraphics[height=0.15\textheight]{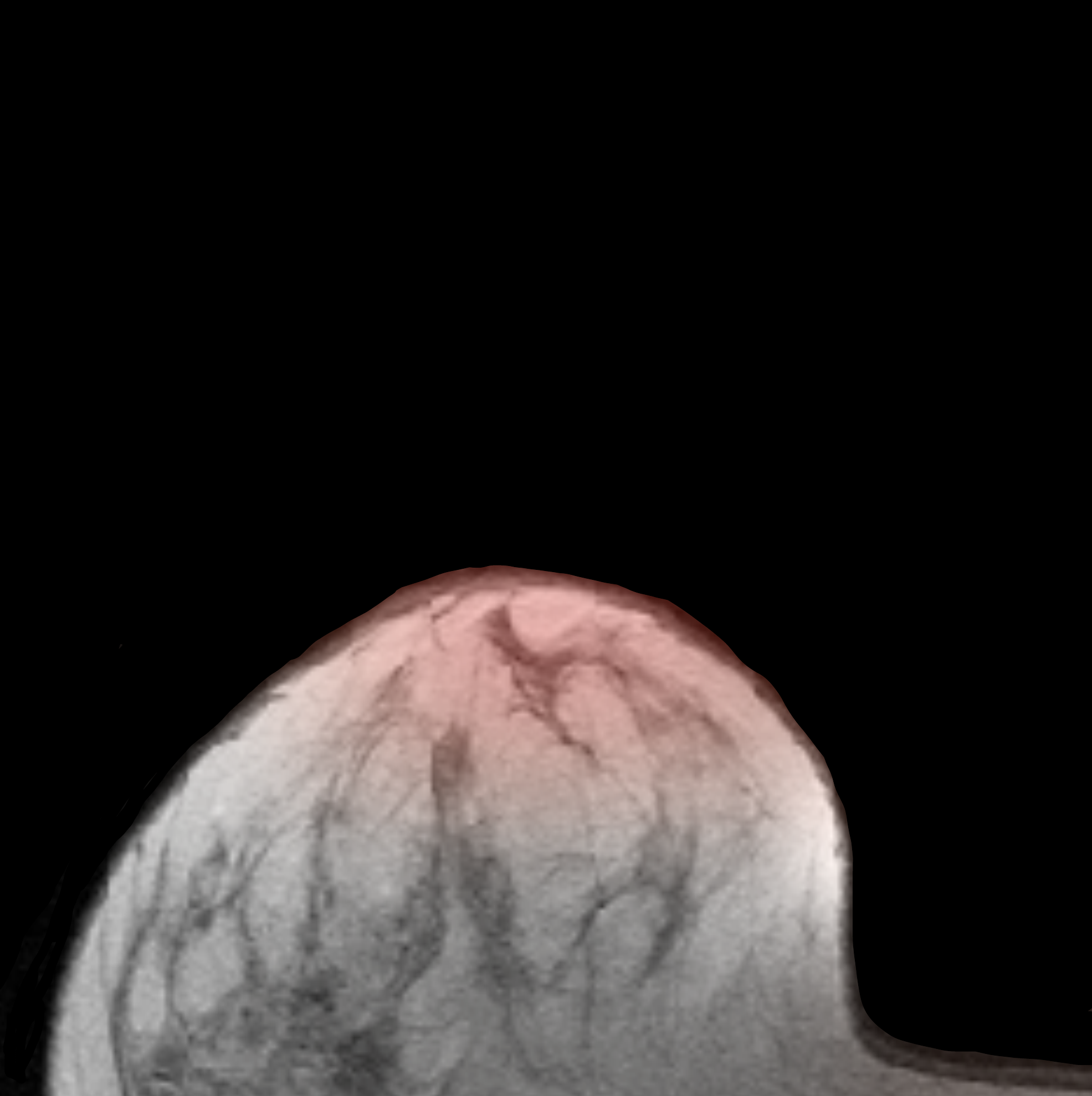}
        % \caption{Malignant, cnn incorrect, rad correct.}
        \caption{}
        \label{fig:sub2}
      \end{subfigure}
    \caption{Class activation maps for (a): Malignant case that was correctly classified by our algorithm and the radiologist. (b): Benign case that was classified correctly by the CNN and incorrectly by the radiologist. (c): Malignant case that was classified incorrectly by both CNN and radiologist. (d): Malignant case that was classified incorrectly by the CNN and correctly by the radiologist. Images have been masked.}
    \label{fig:cm}
\end{figure}

\noindent In Fig.~\ref{fig:cm}, class activation maps for ResNet18 Curriculum are shown.
Regions with high activations for malignancy are marked in red, in order to provide clinicians with enhanced guidance by our algorithm.

\section{Discussion}
Our proposed multi scale curriculum training enabled successful training of the relatively simple 3D ResNet18 from an AUROC of 0.5 to 0.89.
In other experiments that are out of the scope of this work, we found that the particular choice of the backbone architecture had negligible effect on the performance.
On our dataset, the performance of our approach is on par with Retina U-Net~\cite{jaeger_2018} and even exceeds the performance of Mask R-CNN~\cite{he_2017}.
Since we did not tune our models extensively with respect to ensembling, the performance in comparison with the reported results in~\cite{jaeger_2018} seem reasonable to us.
We hypothesize that the high performance of our model at least partly arises from the high amount of context and global information that is provided.
Both Retina U-Net and Mask R-CNN incorporate more model parameters and hyperparameters such as the IoU threshold.
Most importantly, these approaches rely on pixelwise segmentations for all individual lesions.
Our approach on the other hand only needs a coarse localization (i.e.~coordinate) for Stage 1 training and only one global label per breast for Stage 2.
Since in most clinical sites performing breast MRI, a global BIRADS classification per breast is assessed and provided along with the image data, global labels are relatively cheap to generate.
This would allow a Stage 2 training with much more training data and more thorough evaluation with much more test data in the future.
Curriculum learning on a series of tasks as proposed in~\cite{maicas_2018} achieved very similar performance on a breast MRI dataset.
Our approach has some similarities with~\cite{lotter_2017}:
Stage 1 training is very similar, yet performed in 3D in our case.
In Stage 2, Lotter et al.~\cite{lotter_2017} concatenate and pool feature vectors of individual (2D) regions while we perform direct predictions of whole images in 3D.
% Evtl. erwähnen, dass das Modell viel komplizierter zu trainieren ist?

Our work has several limitations:
Firstly, our dataset is monocentric and even though to our best knowledge it is the largest breast MRI dataset for computer aided diagnosis, it is limited in size.
However, extension is relatively easy since only global labels are required for Stage 2 training and assessment of test performance.
For Stage 1 a coarse localization that has to be marked manually is still required.
% Moreover, the class activation maps have several limitations and drawbacks.
In comparison to the diagnostic performance of a breast radiologist, our model is inferior to human performance.

For future work, we will expand the size of our Stage 2 dataset.
Since breast MRI experts are rare, it would be interesting to assess whether our algorithm (including the activation maps) can be applied to teach radiologists with limited experience.
Moreover, it would be interesting to assess whether a combination of an expert rater with our algorithm could further improve the overall performance.

\section{Conclusion}
We presented a simple 3D CNN architecture that is able to perform breast cancer malignancy classification based on magnetic resonance images using a multi scale curriculum learning strategy.
The network architecture predicts malignant breast cancer globally for a whole 3D volume \textit{without} scoring individual lesions.
This approach provides the whole spatial context of a breast to the network, yielding state of the art performance and without the need for lesion segmentations.

\section*{Acknowledgements}
The authors thank Paul Jäger for making Medical Detection Toolkit~\cite{jaeger_2018} publicly available, which was a great help when comparing our method with Mask R-CNN and Retina U-Net.
\bibliographystyle{splncs04}
{\small \bibliography{literature}}
\end{document}